\documentclass[12pt]{article}
\usepackage{amsfonts}
\usepackage{amssymb}
\usepackage{xcolor}
\usepackage{color}
\usepackage[english]{babel}
\usepackage{graphics}
\usepackage{graphicx}
\begin{document}
\vspace{-48pt}
\title{{\bf Long time deviations\\
from the exponential decay law: \\
possible effects \\
in particle physics and cosmology\footnote{Presentation prepared
for  \textbf{XXII--nd Recontres de Blois: Particle Physics and Cosmology,}
Blois, France,  15th -- 20th July 2010.}}}
\author{K. Urbanowski\footnote {e--mail:
K.Urbanowski@proton.if.uz.zgora.pl; K.Urbanowski@if.uz.zgora.pl}, \\
University of Zielona G\'{o}ra, Institute of Physics, \\
ul. Prof. Z. Szafrana 4a, 65--516 Zielona G\'{o}ra, Poland.\\}
\date{}

\maketitle

\begin{abstract}
An effect generated by the nonexponential behavior of the survival
amplitude of an unstable state
in the long time region is considered.
We find that the instantaneous energy of the unstable state for
a large class of models of unstable states
tends to the minimal energy of the system ${\cal E}_{min}$
as $t\rightarrow\infty$ which
is much smaller than the energy of this state for $t$ of the
order of the lifetime of the considered
state. Analyzing the transition time region between exponential
and non-exponential form of the
survival amplitude we find that the instantaneous energy of the
considered unstable state can take
large values, much larger than the energy of this state for $t$
from the exponential time region.
Taking into account results obtained for a model considered,
it is hypothesized that this purely
quantum mechanical effect may be responsible for the properties
of broad resonances such as $\sigma$
meson  as well as having astrophysical and cosmological consequences.
\end{abstract}


\section{Introduction}

Searching for the properties of unstable states $|\phi \rangle
\in {\cal H}$ (where ${\cal H}$ is the Hilbert space of states of
the considered system)
one analyzes their decay law.
The decay law, ${\cal P}_{\phi}(t)$ of
an unstable state $|\phi\rangle$ decaying
in vacuum
is defined as follows
\begin{equation}
{\cal P}_{\phi}(t) = |a(t)|^{2}, \label{P(t)}
\end{equation}
where $a(t)$ is  the probability amplitude of finding the system at the
time $t$ in the initial state $|\phi\rangle$ prepared at time $t_{0}
= 0$,
\begin{equation}
a(t) = \langle \phi|\phi (t) \rangle . \label{a(t)}
\end{equation}
and $|\phi (t)\rangle$ is the solution of the Schr\"{o}dinger equation
for the initial condition  $|\phi (0) \rangle = |\phi\rangle$:
\begin{equation}
i\hbar \frac{\partial}{\partial t} |\phi (t) \rangle = H |\phi (t)\rangle,
\label{Schrod}
\end{equation}
where $H$ denotes the total selfadjoint Hamiltonian for the system
considered.
From basic principles of quantum theory it is known that the
amplitude $a(t)$, and thus the decay law ${\cal P}_{\phi}(t)$ of the
unstable state $|\phi\rangle$, are completely determined by the
density of the energy distribution $\omega({\cal E})$ for the system
in this state \cite{Fock,Fonda}
\begin{equation}
a(t) = \int_{Spec.(H)} \omega({\cal E})\;
e^{\textstyle{-\frac{i}{\hbar}\,{\cal E}\,t}}\,d{\cal E}.
\label{a-spec}
\end{equation}
where $\omega({\cal E}) \geq 0$ and $a(0) = 1$.

In \cite{Khalfin}
assuming that the spectrum of $H$ must be bounded
from below, $(Spec.(H)\; > \; -\infty)$, and using the Paley--Wiener
Theorem  \cite{Paley}
it was proved that in the case of unstable
states there must be $|a(t)| \; \geq \; A\,\exp\,[{\textstyle - b \,t^{q}}] $
for $|t| \rightarrow \infty$ (where $A > 0,\,b> 0$ and $ 0 < q < 1$).

The problem of  how to detect possible deviations from the exponential
form of
${\cal P}_{\phi}(t)$ at the long time region has been attracting
the  attention of physicists since the
first theoretical predictions of such an effect.
Many tests of the decay law performed some time ago
did not indicate
any deviations
from the exponential form of ${\cal P}_{\phi}(t)$ at  the
long time region.
Nevertheless, conditions leading to the nonexponetial behavior
of the amplitude $a(t)$ at long times were studied theoretically.
Conclusions following from these studies were applied successfully
in an experiment described  in the Rothe paper
\cite{rothe}, where the experimental evidence of deviations from
the the exponential decay law at long times was
reported. This result gives rise to another problem which now becomes
important:
If (and how) deviations from the
exponential decay law at long times affect the energy of the unstable state
and its decay rate at this time region.

Note that in fact the amplitude $a(t)$ contains information about
the decay law ${\cal P}_{\phi}(t)$ of the state $|\phi\rangle$, that
is about the decay rate $\gamma_{\phi}^{0}$ of this state, as well
as the energy ${\cal E}_{\phi}^{0}$ of the system in this state.
This information can be extracted from $a(t)$. Indeed, if
$|\phi\rangle$ is an unstable (a quasi--stationary) state then,
there is
\begin{equation}
{\cal E}_{\phi}^{0} - \frac{i}{2} \gamma_{\phi}^{0} \equiv i
\hbar\,\frac{\partial a_{0}(t)}{\partial t} \; \frac{1}{a_{0}(t)},
\label{E-iG}
\end{equation}
for $t \sim \tau_{\phi}$, where
\begin{equation}
a_{0}(t) = \exp\,[- \frac{i}{\hbar}({\cal E}_{\phi}^{0} -
\frac{i}{2} \gamma_{\phi}^{0})t] \simeq a(t), \label{a-q-stat}
\end{equation}
for $t \sim \tau_{\phi}$, $\tau_{\phi} = \frac{\hbar}{\gamma^{0}_{\phi}}$
and   $\gamma^{0}_{\phi}$ is the decay rate of $|\phi\rangle$.

The standard interpretation and understanding of the quantum theory
and the related construction of our measuring devices are such that
detecting the energy ${\cal E}_{\phi}^{0}$ and decay rate
$\gamma_{\phi}^{0}$ one is sure that the amplitude $a(t)$ has the
form (\ref{a-q-stat}) and thus that the relation (\ref{E-iG})
occurs.
Taking the above into account one can define the "effective
Hamiltonian", $h_{\phi}$, for the one--dimensional subspace of
states ${\cal H}_{||}$ spanned by the normalized vector
$|\phi\rangle$ as follows  (see, eg. \cite{PRA}),
\begin{equation}
h_{\phi} \stackrel{\rm def}{=}  i \hbar\, \frac{\partial
a(t)}{\partial t} \; \frac{1}{a(t)}. \label{h}
\end{equation}
In general, $h_{\phi}$ can depend on time $t$, $h_{\phi}\equiv
h_{\phi}(t)$.

It is easy to show that equivalently \cite{urbanowski-2-2009}
\begin{equation}
h_{\phi} (t) \equiv
\frac{\langle \phi|H|\phi (t)\rangle}{\langle \phi |\phi (t)\rangle}.
\label{h-equiv}
\end{equation}
One meets  effective Hamiltonians of this type when one starts
with the time--depen\-dent Schr\"{o}dinger equation (\ref{Schrod})
for the total state
space ${\cal H}$ and looks for the rigorous evolution equation for
the distinguished subspace of states ${\cal H}_{||} \subset {\cal H}$
(see \cite{PRA} and references one finds therein). In the case of
one--dimensional ${\cal H}_{||}$  this rigorous
Schr\"{o}dinger--like evolution equation has the following form for
the initial condition $a(0) = 1$, \cite{PRA},
\begin{equation}
i \hbar\, \frac{\partial a(t)}{\partial t} \;=\; h_{\phi}(t)\;a(t).
\label{eq-for-h}
\end{equation}
Relations (\ref{h}) and (\ref{eq-for-h}) establish a direct
connection between the amplitude $a(t)$ for the state $|\phi
\rangle$ and the exact effective Hamiltonian $h_{\phi}(t)$ governing
the time evolution in the one--dimensional subspace ${\cal H}_{\|}
\ni |\phi\rangle$. Thus, the use of the evolution equation
(\ref{eq-for-h}) or the relation (\ref{h}) is one of the most
effective tools for the accurate analysis of the early-- as well as
the long--time properties of the energy and decay rate of a given
quasi--stationary state $|\phi (t) \rangle$.

So let us assume that we know the amplitude $a(t)$. Then starting
with this $a(t)$ and using the expression (\ref{h}) one can
calculate the effective Hamiltonian $h_{\phi}(t)$ in a general case
for every $t$. Thus, one finds the following expressions for the
energy and the decay rate of the system in the state $|\phi\rangle$
under considerations, to be more precise for the instantaneous
energy and the instantaneous decay rate,
(for details see: \cite{urbanowski-2-2009}),
\begin{eqnarray}
{\cal E}_{\phi}&\equiv& {\cal E}_{\phi}(t) = \Re\,(h_{\phi}(t),
\label{E(t)}\\
\gamma_{\phi} &\equiv& \gamma_{\phi}(t) = -\,2\,\Im\,(h_{\phi}(t),
\label{G(t)}
\end{eqnarray}
where $\Re\,(z)$ and $\Im\,(z)$ denote the real and imaginary parts
of $z$ respectively.

Using (\ref{h}) and (\ref{E(t)}), (\ref{G(t)}) one can find that
\begin{eqnarray}
{\cal E}_{\phi} (0) &=& \langle \phi |H| \phi \rangle, \\
{\cal E}_{\phi} (t \sim \tau_{\phi}) & \simeq &
{\cal E}_{\phi}^{0} \;\;\neq \;\; {\cal E}_{\phi} (0),\\
\gamma_{\phi}(0) &=& 0,\\
\gamma_{\phi}(t \sim \tau_{\phi}) &\simeq & \gamma_{\phi}^{0}.
\end{eqnarray}

The aim of this talk is to discuss the long time behaviour of ${\cal
E}_{\phi}(t)$  using $a(t)$ calculated for the given density
$\omega({\cal E})$. We show that ${\cal E}_{\phi}(t)
\rightarrow {\cal E}_{min} > - \infty$
as $t\rightarrow \infty$ for the model considered and that a wide
class of models has similar long time properties:
${{{\cal E}_{\phi}(t)}\vline}_{\;t \rightarrow \infty}
\neq {\cal E}_{\phi}^{0}$.
It seems that, in contrast to the standard Khalfin
effect \cite{Khalfin},
in the case of the quasi--stationary states
belonging to the same class as excited atomic levels,  these long time
properties of the instantaneous energy ${\cal E}_{\phi}(t)$ have
a chance to be
detected, eg.,  by analyzing the properties of the high energy
cosmic rays  or the spectra of very distant astrophysical objects.

\section{The model}

Let us assume that ${Spec. (H)} = [{\cal E}_{min}, \infty)$,
(where, ${\cal E}_{min} > - \infty$), and let us choose
$\omega ({\cal E})$ as follows (compare \cite{Sluis})
\begin{equation}
\omega ({\cal E}) \equiv \omega_{BW}({\cal E}, {\cal E}_{min}) =
\frac{N}{2\pi}\,  \it\Theta ({\cal E} - {\cal E}_{min}) \
\frac{\gamma_{\phi}^{0}}{({\cal E}-{\cal E}_{\phi}^{0})^{2} +
(\frac{\gamma_{\phi}^{0}}{2})^{2}}, \label{omega-BW}
\end{equation}
where $N$ is a normalization constant and
\[\it\Theta ({\cal E}) \ = \left\{
  \begin{array}{c}
   1 \;\;{\rm for}\;\; {\cal E} \geq 0, \\
   0 \;\; {\rm for}\;\; {\cal E} < 0.\\
  \end{array}
\right.
\]
For such $\omega_{BW}({\cal E})$ using (\ref{a-spec}) one has
\begin{equation}
a(t) = \frac{N}{2\pi}  \int_{{\cal E}_{min}}^{\infty}
 \frac{{\gamma_{\phi}^{0}}}{({\cal E}-{\cal E}_{\phi}^{0})^{2}
+ (\frac{\gamma_{\phi}^{0}}{2})^{2}}\, e^{\textstyle{ -
\frac{i}{\hbar}{\cal E}t}}\,d{\cal E}, \label{a-BW}
\end{equation}
where
\begin{equation}
\frac{1}{N} = \frac{1}{2\pi} \int_{{\cal E}_{min}}^{\infty}
 \frac{\gamma_{\phi}^{0}}{({\cal E}-{\cal E}_{\phi}^{0})^{2}
+ (\frac{\gamma_{\phi}^{0}}{2})^{2}}\, d{\cal E}. \label{N}
\end{equation}
Formula  (\ref{a-BW}) leads to the result (see also \cite{Sluis})
\begin{eqnarray}
a(t) &=& N\,e^{\textstyle{- \frac{i}{\hbar} ({\cal
E}_{\phi}^{0} -
i\frac{\gamma_{\phi}^{0}}{2})t}}\,
\Big\{1 - \frac{i}{2\pi} \times\nonumber \\
&& \times\,\Big[
e^{\textstyle{\frac{\gamma_{\phi}^{0}t}{\hbar}}}\,
E_{1}\Big(-\frac{i}{\hbar}({\cal E}_{\phi}^{0} -{\cal E}_{min}
+ \frac{i}{2} \gamma_{\phi}^{0})t\Big) \nonumber\\
&&\,-\, E_{1}\Big(- \frac{i}{\hbar}({\cal E}_{\phi}^{0} -{\cal E}_{min} -
\frac{i}{2} \gamma_{\phi}^{0})t\Big)\,\Big]\, \Big\}, \label{a-E(1)}
\end{eqnarray}
where $E_{1}(x)$ denotes the integral--exponential function
\cite{Sluis,Abramowitz}.

In general one has
\begin{equation}
a(t) \equiv a_{exp}(t) + a_{non}(t),
\label{a-exp+a-non}
\end{equation}
where
\[
a_{exp}(t) = N\,e^{\textstyle{- \frac{i}{\hbar} ({\cal
E}_{\phi}^{0} -
i\frac{\gamma_{\phi}^{0}}{2})t}}, \;\;\;\;\;\;\;\; a_{non}(t) =
a(t) - a_{exp}(t).
\]

Making use of  the asymptotic expansion of $E_{1}(x)$ \cite{Abramowitz}
\begin{equation}
{E_{1}(z)\vline}_{\, |z| \rightarrow \infty} \;\;\sim \;\;
\frac{e^{\textstyle{ -z}}}{z}\,( 1 - \frac{1}{z} + \frac{2}{z^{2}} -
\ldots ),  \label{E1-as}
\end{equation}
where $| \arg z  | < \frac{3}{2} \pi$, one finds
\begin{eqnarray}
{a(t)\vline}_{\, t \rightarrow \infty} &\simeq & N
e^{\textstyle - \frac{i}{\hbar}\,h_{\phi}^{0}\,t}
\nonumber \\&&
\;+\;
\frac{N}{2 \pi}\;e^{\textstyle{-\frac{i}{\hbar}\,{\cal E}_{min}t}}\;
\Big\{
(- i)\; \frac{
\gamma_{\phi}^{0}}{|\,h_{\phi}^{0}-{\cal E}_{min}\,|^{\,2}}
 \, \frac{\hbar}{t}  \nonumber \\
&&\,- 2\,\frac{({\cal E}_{\phi}^{0}\,-\,{\cal E}_{min})\,
\gamma_{\phi}^{0}}{|\,h_{\phi}^{0}\,-\,{\cal E}_{min}\,|^{\,4}} \,
\Big(\frac{\hbar}{t}\Big)^{2}\,+ \ldots\Big\} \label{a(t)-as}
\end{eqnarray}
where $h^{0}_{\phi} = {\cal E}_{\phi}^{0} \,-
\,\frac{i}{2}\,\gamma_{\phi}^{0}$,  and
\begin{eqnarray}
{h_{\phi}(t)\vline}_{\,t \rightarrow \infty} & = &
{i \hbar \,\frac{\partial a(t)}{\partial
t}\,\frac{1}{a(t)} \vline}_{\,t \rightarrow \infty} \nonumber \\
&\simeq & {\cal E}_{min}\,
-\,i\,\frac{\hbar}{t}\;  - \;2\,
\frac{ {\cal E}_{\phi}^{0}\,-\,{\cal E}_{min}}{|\,h_{\phi}^{0}\,-
\,{\cal E}_{min} \,|^{\,2} }  \; \Big( \frac{\hbar}{t} \Big)^{2}
\;+\ldots \;\; \label{h-as}
\end{eqnarray}
for the considered case (\ref{omega-BW}) of $\omega_{BW}({\cal E})$
(for details see \cite{urbanowski-2-2009}).
From (\ref{h-as}) it follows that
\begin{eqnarray}
\Re\,({h_{\phi}(t)\vline}_{\,t \rightarrow \infty}) &\stackrel{\rm
def}{=}& {\cal E}_{\phi}^{\infty} (t)\, \nonumber \\
&\simeq & {\cal E}_{min}
 -\,2\,
\frac{ {\cal E}_{\phi}^{0}\,-
\,{\cal E}_{min}}{ |\,h_{\phi}^{0}\,-\,{\cal E}_{min} \,|^{\,2} }
\; \Big(
\frac{\hbar}{t} \Big)^{2} \nonumber \\ &&
                            \begin{array}{c}
                               {} \\
                               \longrightarrow \\
                               \scriptstyle{t \rightarrow \infty}
                             \end{array}
                             \, {\cal E}_{min},\label{Re-h-as}
\end{eqnarray}
where ${\cal E}_{\phi}^{\infty}(t) = {\cal
E}_{\phi}(t)|_{\,t \rightarrow \infty}$,
and
\begin{equation}
\Im\,({h_{\phi}(t)\vline}_{\,t \rightarrow \infty}) \simeq
-\,\frac{\hbar}{t} \,
\begin{array}{c}
                               {} \\
                               \longrightarrow \\
                               \scriptstyle{t \rightarrow \infty}
                             \end{array}
\,0. \label{Im-h-as}
\end{equation}
The property (\ref{Re-h-as}) means that
\begin{equation}
\Re\,({h_{\phi}(t)\vline}_{\,t \rightarrow \infty})\,\equiv\,
{\cal E}_{\phi}^{\infty}(t)\,
<\, {\cal E}_{\phi}^{0}. \label{E-infty<E0}
\end{equation}
For different states $|\phi\rangle \, =\,|j\rangle$, ($j = 1,2,3,\ldots$)
one has
\begin{equation}
\Im\,({h_{1}(t)\vline}_{\,t \rightarrow \infty}) =
\Im\,({h_{2}(t)\vline}_{\,t \rightarrow
\infty}),\label{Im-h1-h2-as}
\end{equation}
whereas in general $\gamma_{1}^{0}\, \neq \,\gamma_{2}^{0}$.

Note that
from (\ref{a(t)-as}) one obtains
\begin{eqnarray}
{\vline\,{a(t)\vline}_{\, t \rightarrow
\infty}\,\vline}^{\,2} &\simeq& N^{2} e^{\textstyle -
\frac{\gamma_{\phi}^{0}}{\hbar}\,\,t} \nonumber \\
&& + \frac{N^{2}}{\pi}\,\,\,\sin\,[({\cal E}_{\phi}^{0} -
{\cal E}_{min})\,t]\,\,\,\,\,
e^{\textstyle - \frac{1}{2}\,\frac{\gamma_{\phi}^{0}}{\hbar}\,\,t}\;\;
\frac{\gamma_{\phi}^{0}}{|h_{\phi}^{0}\,-\,{\cal E}_{min}\,|^{\,2}} \;
\frac{\hbar}{t} \nonumber \\
&& + \frac{N^{2}}{4
\pi^{2}}\;\frac{(\gamma_{\phi}^{0})^{2}}{|h_{\phi}^{0}\,-
\,{\cal E}_{min}\,|^{\,4}} \;
\frac{\hbar^{2}}{t^{2}}\; + \;\ldots\;\; . \label{t-as-1}
\end{eqnarray}
Relations (\ref{a(t)-as}) ---  (\ref{Im-h1-h2-as})
become important
for times $t > t_{as}$, where $t_{as}$ denotes the time $t$ at which
contributions to ${\vline\,{a(t)\vline}_{\, t \rightarrow
\infty}\,\vline}^{\,2}$ from the first exponential component in
(\ref{t-as-1}) and from the third component proportional to
$\frac{1}{t^{2}}$ are comparable, that is (see (\ref{a-exp+a-non})),
\begin{equation}
|a_{exp}(t)|^{2} \,\simeq\,|a_{non}(t)|^{2}
\label{a-exp=a-non}
\end{equation}
for $t\rightarrow \infty $.
So $t_{as}$ can be be found by
considering the following relation
\begin{equation}
e^{\textstyle - \frac{ \gamma_{\phi}^{0}}{\hbar}\,\,t}\; \sim\;
\frac{1}{4
\pi^{2}}\;\frac{(\gamma_{\phi}^{0})^{2}}{|h_{\phi}^{0}\,-
\,{\cal E}_{min}\,|^{\,4}} \;
\frac{\hbar^{2}}{t^{2}}. \label{t-as-2}
\end{equation}
Assuming that the right hand side is equal to the left hand side in
the above relation one gets a transcendental equation.  Exact
solutions of such an equation can be expressed by means of the
Lambert $W$ function \cite{Corless}.
An asymptotic solution of the
equation obtained from the relation (\ref{t-as-2}) is relatively
easy to find \cite{Olver}.
The very approximate asymptotic solution,
$t_{as}$, of this equation for $(\frac{{\cal
E}_{\phi}}{\gamma_{\phi}^{0}})\,>\,10$ (in general for
$(\frac{{\cal E}_{\phi}}{\gamma_{\phi}^{0}})\,\rightarrow \,\infty$)
has the form
\begin{eqnarray}
\frac{\gamma_{\phi}^{0}\,t_{as}}{\hbar} &\simeq & 8,28 \,+\, 4\,
\ln\,(\frac{{\cal E}_{\phi}^{0}\,-
\,{\cal E}_{min}}{\gamma_{\phi}^{0}}) \nonumber \\
&&+\, 2\,\ln\,[8,28 \,+\,4\,\ln\,(\frac{{\cal
E}_{\phi}^{0}\,-\,{\cal E}_{min}}{\gamma_{\phi}^{0}})\,]\,+\, \ldots \;\;.
\label{t-as-3}
\end{eqnarray}

\section{Some generalizations}

To complete the analysis performed in the previous Section let us
consider a more general case of $\omega ({\cal E})$ and $a(t)$.
For a start, let us consider a   relatively simple case when
$\lim_{{\cal E} \rightarrow {\cal E}_{min}+}
\;\omega ({\cal E})\stackrel{\rm def}{=}
\omega_{0}>0$ and ${\omega ({\cal E})\vline}_{\;{\cal E}\, <
\,{\cal E}_{min}}\,=\,0$.   Let  derivatives  $\omega^{(k)}({\cal E})$,
($k= 0,1,2, \ldots, n$),  be continuous
in  $[ {\cal E}_{min}, \infty)$, (that is let for ${\cal E} >
{\cal E}_{min}$ all
$\omega^{(k)}({\cal E})$ be continuous
and all the limits
$\lim_{{\cal E} \rightarrow {\cal E}_{min}+}\,\omega^{(k)}({\cal E})$
exist)  and
let all these $\omega^{(k)}({\cal E})$ be absolutely integrable functions then
(see \cite{urbanowski-1-2009}),
\begin{equation}
a(t) \; \begin{array}{c}
          {} \\
          \sim \\
          \scriptstyle{t \rightarrow \infty}
        \end{array}
        \;- \frac{i\hbar}{t}\;
        e^{\textstyle{-\frac{i}{\hbar}{\cal E}_{min} t}}\;
        \sum_{k = 0}^{n-1}(-1)^{k} \,
        \big(\frac{i\hbar}{t}\big)^{k}\,\omega^{(k)}_{0},
        \label{a-omega}
\end{equation}
where $\omega^{(k)}_{0}  \stackrel{\rm def}{=}
\lim_{{\cal E}\rightarrow {\cal E}_{min}+}
\;\omega^{(k)} ({\cal E})$.

Let us now consider a more complicated form of the density
$\omega ({\cal E})$. Namely
let $\omega ({\cal E})$ be of the form
\begin{equation}
\omega ({\cal E}) = ( {\cal E} - {\cal E}_{min})^{\lambda}\;
\eta ({\cal E})\; \in \; L_{1}(-\infty, \infty),
\label{omega-eta}
\end{equation}
where $0 < \lambda < 1$ and it is assumed that $\eta ({\cal E})
\geq 0$, $\eta^{(k)}({\cal E})$,
($k= 1, 2, \ldots, n$), \linebreak exist and they are continuous
in $[{\cal E}_{min}, \infty)$, and  limits
$\lim_{{\cal E} \rightarrow {\cal E}_{min}+}\;\eta^{(k)}({\cal E})$ exist,
$\lim_{{\cal E} \rightarrow \infty}\;( {\cal E} -
{\cal E}_{min})^{\lambda}\,\eta^{(k)}({\cal E}) = 0$
for all above mentioned $k$ and \linebreak
${\omega ({\cal E})\vline}_{\;{\cal E}\, < \,{\cal E}_{min}}\,=\,0$, then
\begin{eqnarray}
a(t) & \begin{array}{c}
          {} \\
          \sim \\
          \scriptstyle{t \rightarrow \infty}
        \end{array} &
        - \frac{i\hbar}{t}\;\lambda\;
        e^{\textstyle{-\frac{i}{\hbar}{\cal E}_{min} t}}\;
        \Big[\alpha_{n}(t) +
        \big(-\,\frac{i\hbar}{t}\,\big)\,\alpha_{n-1}(t) \nonumber \\
        && + \big(-\,\frac{i\hbar}{t}\big)^{2}\,\alpha_{n-2}(t)
        \nonumber \\
        && + \big(-\,\frac{i\hbar}{t}\big)^{3}\,
        \alpha_{n-3}(t)+ \dots \Big],
        \label{a-eta}
\end{eqnarray}
where (compare \cite{erdelyi,copson})
\begin{equation}
\alpha_{n-k}(t) = \sum_{l=0}^{n-k-1}\,
\frac{\Gamma (l + \lambda)}{l!}\;\,e^{\textstyle{-\,i\,
\frac{\pi (l + \lambda +2)}{2}}}
\;\eta_{0}^{(l + k)}\,\big(\frac{\hbar}{t} \big)^{l + \lambda}, \label{alpha}
\end{equation}
$\Gamma (z)$ is the Gamma Function and $\eta^{(j)}_{0}  =
\lim_{{\cal E} \rightarrow {\cal E}_{min}+}
\;\eta^{(j)} ({\cal E})$, $\eta^{(0)}({\cal E}) =
\eta ({\cal E})$ and $j = 0,1, \ldots,n$.

The asymptotic form of $h_{u}(t)$ for $t \rightarrow \infty$ for
the $a(t)$ given by the relation  (\ref{a-omega}) looks as follows
\begin{eqnarray}
h_{u}^{\infty}(t)\; \stackrel{\rm def}{=}\;
         {h_{u}(t)\,\vline}_{\;t \rightarrow \infty} \;
         &=& \;{\cal E}_{min} \,-\,i\,\frac{\hbar}{t}\nonumber \\
         &&-\frac{\omega_{0}^{(1)}}{\omega_{0}}\;
         \big(\,\frac{\hbar}{t}\,\big)^{2}\;+\;\ldots\;\, . \label{h-infty}
\end{eqnarray}

In the more general case  of $a(t)$ (see, e.g. (\ref{a-eta}) )
after some algebra the
asymptotic approximation  of $a(t)$  can be written as follows
\begin{equation}
a(t) \;\;
\begin{array}{c}
   {} \\
   \sim\\
   {\scriptstyle t \rightarrow \infty}
 \end{array}
\;\;e^{\textstyle{-i\frac{t}{\hbar}\,{\cal E}_{min}}}\;
\sum_{k=0}^{N} \,\frac{c_{k}}{t^{\xi + k}},
\label{a(t)-as-w}
\end{equation}
where $\xi \geq 0$ and $c_{k}$ are complex numbers.

From the relation (\ref{a(t)-as-w})
one concludes that
\begin{equation}
\frac{\partial a(t)}{\partial t} \;\;
\begin{array}{c}
   {} \\
   \sim\\
   {\scriptstyle t \rightarrow \infty}
 \end{array}
\;\; e^{\textstyle{-i\frac{t}{\hbar}\,{\cal E}_{min}}}\;
\Big\{-\frac{i}{\hbar}\, {\cal E}_{min}\,
-\,\sum_{k=0}^{N} \,(\xi + k)\,\frac{c_{k}}{t^{\xi + k
+ 1}}\Big\}. \label{da(t)-as-w}
\end{equation}

Now let us take into account the relation (\ref{eq-for-h}). From
this relation and relations (\ref{a(t)-as-w}), (\ref{da(t)-as-w}) it
follows that
\begin{equation}
h_{\phi}(t) \;\;
\begin{array}{c}
   {} \\
   \sim\\
   {\scriptstyle t \rightarrow \infty}
 \end{array}
\;\;
{\cal E}_{min}\,+\,
\frac{d_{1}}{t}\, +
\,\frac{d_{2}}{t^{2}}\,+\,\frac{d_{3}}{t^{3}}\,+\,\ldots\,\,\, ,
\label{h-sim-w}
\end{equation}
where $d_{1}, d_{2}, d_{3}, \ldots$ are complex numbers
with negative or positive real and imaginary parts. This means
that in the case of the asymptotic approximation to $a(t)$ of
the form (\ref{a(t)-as-w}) the following  property holds,
\begin{equation}
\lim_{t \rightarrow \infty}\,h_{\phi}(t) \, =
\, {\cal E}_{min}\, <\, {\cal E}_{\phi}^{0}. \label{lim-h}
\end{equation}

It seems to be important that  results (\ref{h-sim-w}) and
(\ref{lim-h}) coincide with the results (\ref{h-as}) ---
(\ref{Im-h1-h2-as})
obtained for the density $\omega_{BW}({\cal E})$
given by the formula (\ref{omega-BW}). This means that general
conclusion obtained for the other $\omega ({\cal E})$ defining
unstable states should be similar to those following from
(\ref{h-as}) --- (\ref{Im-h1-h2-as}).

\section{Numerical calculations}

Long time properties of the survival probability $|a(t)|^{2}$
and instantaneous energy ${\cal E}_{\phi}(t)$ are relatively
easy to find analytically for times $t \gg t_{as}$ even in
the general case as it was shown in previous Section and
\cite{urbanowski-1-2009}. It is much more difficult to analyze
these properties analytically in the transition time region
where $t \sim t_{as}$. It can be done numerically for given
models (see \cite{U+P-2009}).

The model considered in Sec. 2 and defined by the density
$\omega_{BW}({\cal E})$, (\ref{omega-BW}), allows one to
find numerically the decay curves and the instantaneous
energy $\varepsilon_{\phi}(t)$ as a function of time $t$.
The results presented in this Section have been obtained
assuming for simplicity that the minimal energy ${\cal E}_{min}$
appearing in the formula (\ref{omega-BW}) is equal to zero,
${\cal E}_{min} = 0$. So, all numerical calculations were
performed for the density $\tilde{\omega}_{BW}({\cal E})$
given by the following formula
\begin{equation}
\tilde{\omega}_{BW}({\cal E}) \equiv
\omega_{BW} ({\cal E},{\cal E}_{min}=0) =
\frac{N}{2\pi}\,  \it\Theta ({\cal E}) \
\frac{\gamma_{\phi}^{0}}{({\cal E}-{\cal E}_{\phi}^{0})^{2} +
(\frac{\gamma_{\phi}^{0}}{2})^{2}}, \label{omega-BW-E-min=0}
\end{equation}
for some chosen $\frac{{\cal E}_{\phi}^{0}}{\gamma_{\phi}^{0}}$.
Performing calculations  particular attention was paid to
the form of the probability $|a(t)|^{2}$, i. e. of the decay
curve, and of the instantaneous energy $\varepsilon_{\phi}(t)$
for times $t$ belonging to the most interesting transition
time-region between exponential and nonexponential parts of
$|a(t)|^{2}$, where the following relation corresponding
with (\ref{a-exp=a-non}) and (\ref{t-as-2}) takes place,
\begin{equation}
|a_{exp}(t)|^{2} \;\sim \;  |a_{non}(t)|^{2},
\label{a-exp=a-non-1}
\end{equation}
where $a_{exp}(t), a_{non}(t)$ are defined by
(\ref{a-exp+a-non}). Results are presented graphically
below in Figs (\ref{a10-1}) --- (\ref{h100}).

\begin{figure}[h]

{\includegraphics[height=60mm,width=110mm]{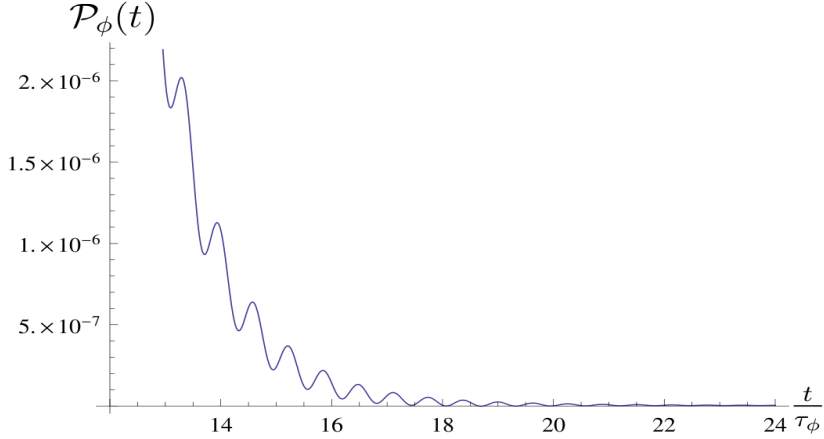}}\\
\caption{Survival probability ${\cal P}_{\phi}(t) = |a(t)|^{2}$
in the transition time region. The case
$\frac{{\cal E}_{\phi}^{0}}{\gamma_{\phi}^{0}} = 10$.}
\label{a10-1}
\vspace*{-12pt}
\end{figure}

\begin{figure}
\begin{center}
\includegraphics[height=60mm,width=110mm]{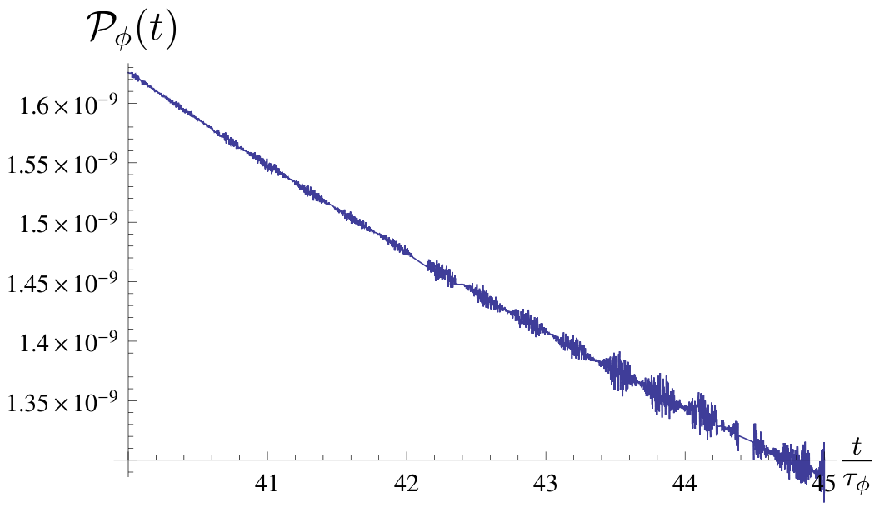}\\
\caption{Survival probability ${\cal P}_{\phi}(t) = |a(t)|^{2}$
in the transition time region.
The case $\frac{{\cal E}_{\phi}^{0}}{\gamma_{\phi}^{0}} = 10$.}
\label{a10-2}
\end{center}
\end{figure}

\begin{figure}
  \includegraphics[width=120mm]{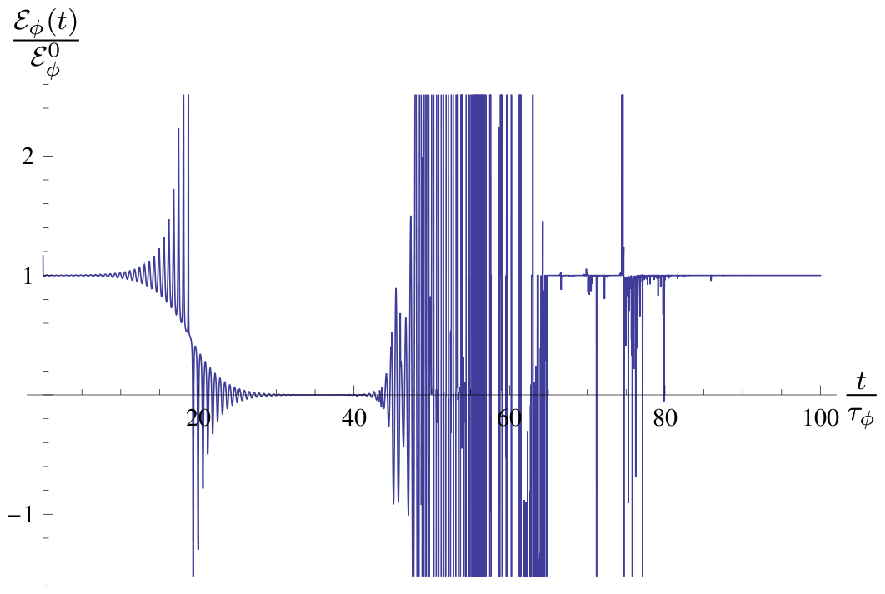}\\
  \caption{Instantaneous energy ${\cal E}_{\phi}(t)$
  in the transition time region. The case
  $\frac{{\cal E}_{\phi}^{0}}{\gamma_{\phi}^{0}} = 10$.}
  \label{h10}
\end{figure}

\begin{figure}
\begin{center}
\includegraphics[height=60mm,width=110mm]{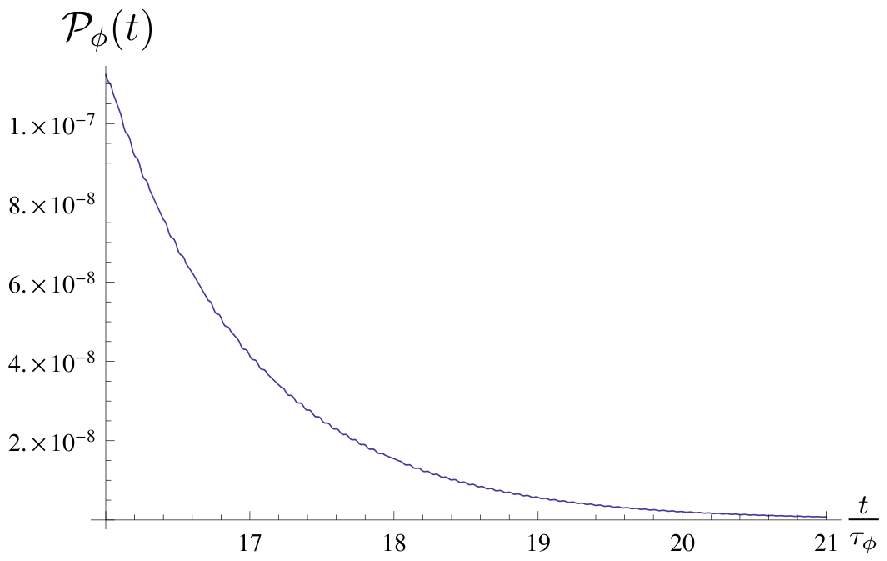}\\
\caption{Survival probability ${\cal P}_{\phi}(t) =
|a(t)|^{2}$ in the transition time region.
The case $\frac{{\cal E}_{\phi}^{0}}{\gamma_{\phi}^{0}} = 100$.}
\label{a100-1}
\end{center}
\end{figure}

\begin{figure}
\begin{center}
\includegraphics[height=60mm,width=110mm]{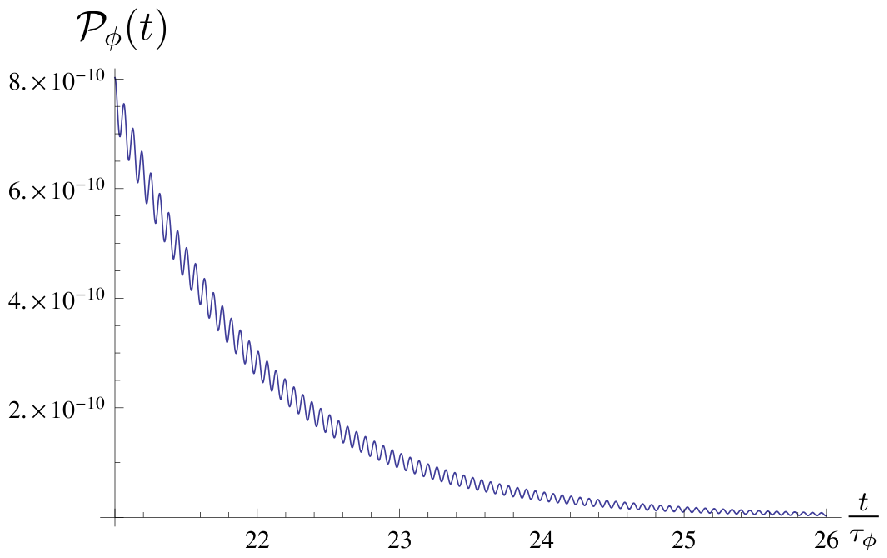}\\
\caption{Survival probability ${\cal P}_{\phi}(t) =
|a(t)|^{2}$ in the transition time region.
The case $\frac{{\cal E}_{\phi}^{0}}{\gamma_{\phi}^{0}} = 100$.}
\label{a100-2}
\end{center}
\end{figure}

\begin{figure}
\begin{center}
\includegraphics[height=60mm,width=110mm]{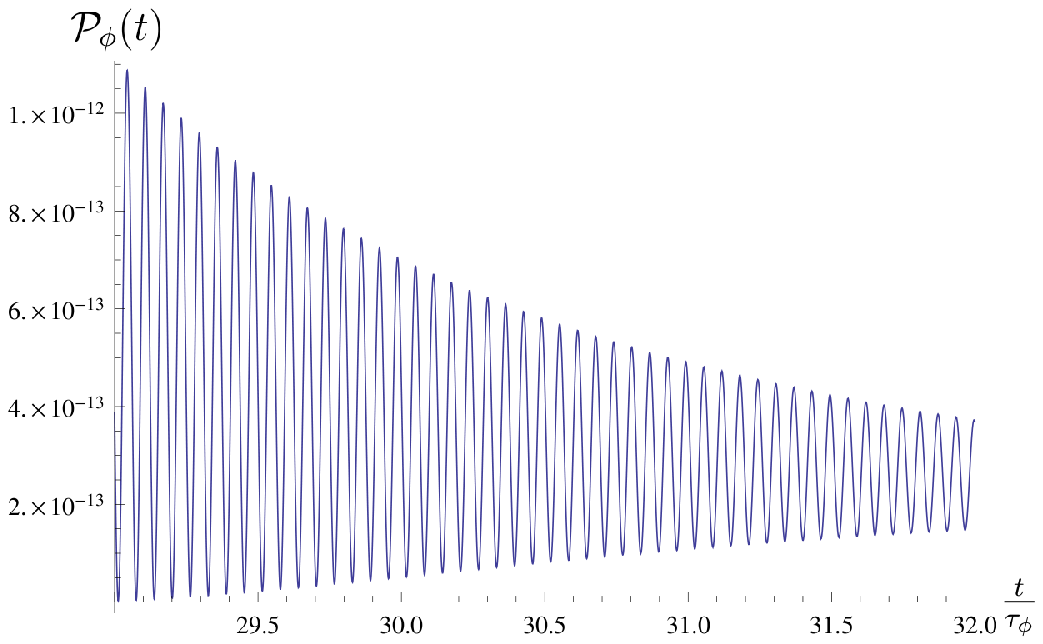}\\
\caption{Survival probability ${\cal P}_{\phi}(t) =
|a(t)|^{2}$ in the transition time region.
The case $\frac{{\cal E}_{\phi}^{0}}{\gamma_{\phi}^{0}} = 100$.}
\label{a100-3}
\end{center}
\end{figure}

\begin{figure}
  \includegraphics[width=120mm]{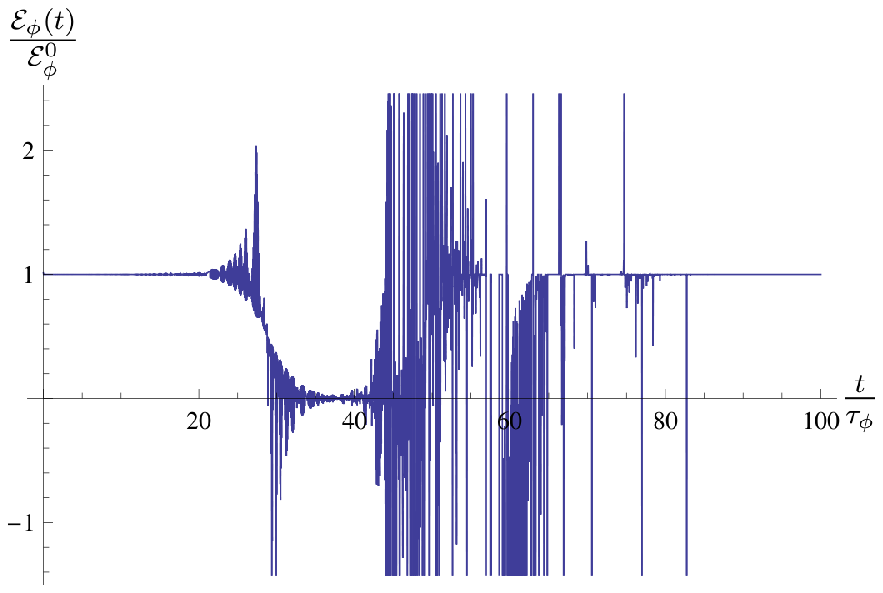}\\
  \caption{Instantaneous energy ${\cal E}_{\phi}(t)$ in the
  transition time region. The case
  $\frac{{\cal E}_{\phi}^{0}}{\gamma_{\phi}^{0}} = 100$.}
  \label{h100}
\end{figure}

\section{Final remarks.}

Decay curves of a type Fig. (\ref{a10-1}), Fig. (\ref{a10-2}),
Fig. (\ref{a100-1}) --- Fig. (\ref{a100-3}) one meets for a very
large class of models defined  by energy densities $\omega({\cal E })$
of the following type (see \cite{Fonda,nowakowski-3}),
\begin{eqnarray}
\omega({\cal E})\,&=&\, \frac{N}{2\pi}\;
{\mathit \Theta}({\cal E} - {\cal E}_{min} )\;({\cal E} -
{\cal E}_{min})^{\lambda} \,
\frac{\gamma_{\phi}^{0}}{({\cal E}- {\cal E}_{\phi}^{0})^{2}+
\frac{(\gamma_{\phi}^{0})^{2}}{4}}\, f({\cal E}),\label{omega-gen}\\
&\equiv & \omega_{BW}({\cal E}, {\cal E}_{min})\;({\cal E} -
{\cal E}_{min})^{\lambda}\, f({\cal E}), \nonumber
\end{eqnarray}
where $\lambda \geq 0$, $f({\cal E})$ is a form--factor ---  it
is a smooth function going to zero as ${\cal E} \rightarrow \infty$
and it has no threshold and no pole. The asymptotical large time
behavior of $a(t)$ is due to the term
$({\cal E} - {\cal E}_{min})^{\lambda}$ and the choice of
$\lambda$ (see Sec. 3). The density $\omega({\cal E })$ defined
by the relation (\ref{omega-gen}) fulfills all physical
requirements and it leads to the decay curves having a very
similar form at transition times region to the decay curves
presented above. The characteristic  feature of all these decay
curves is the presence of sharp and frequent oscillations at
the transition times region (see Figs (\ref{a10-1}), (\ref{a10-2}),
(\ref{a100-1}), (\ref{a100-2}), (\ref{a100-3}))
(see also, eg. \cite{calderon1,calderon2}). This means that
derivatives of the amplitude $a(t)$ may reach extremely large
values for some times from the transition time region and the
modulus of these derivatives is much larger than the modulus
of $a(t)$, which is very small for these times. This explains
why in this time region the real and imaginary parts of $h_{\phi}(t)
\equiv {\cal E}_{\phi}(t) \, - \,\frac{i}{2}\,\gamma_{\phi}(t)$,
which can be expressed by the relation (\ref{h}), ie. by a large
derivative of $a(t)$  divided by a very small $a(t)$, reach
values much larger than the energy ${\cal E}_{\phi}^{0}$ of
the the unstable state measured at times for which the decay
curve has the exponential form. For the model considered we
found that, eg. for $\frac{{\cal E}_{\phi}^{0}}{\gamma_{\phi}^{0}}
= 10$ and   $5 \tau_{\phi}\, \leq \,t\,\leq\,60 \tau_{\phi}$
the maximal value of the instantaneous energy equals
${\cal E}_{\phi}(t)\, = \,89,2209 \,{\cal E}_{\phi}^{0}$ and
${\cal E}_{\phi}(t)$ reaches this value for
$t\equiv t_{mx,\;10} = 53,94\,\tau_{\phi}$   and then the survival
probability ${\cal P}_{\phi}(t)$ is of order
${\cal P}_{\phi}(t_{mx,\,10}) \sim 10^{-9}$.

The question is whether and where this effect can manifest
itself. There are two possibilities to observe the above long
time properties of unstable states: The first one is that one
should analyze properties of unstable states having not too
long values of $t_{as}$. The second one is finding a possibility
to observe a suitably large number of events, i.e. unstable
particles, created by the same source.

The problem with understanding the properties of broad
resonances in the scalar sector ($\sigma$ meson
problem \cite{pdg-2008}) discussed in \cite{nowakowski-1,nowakowski-2},
where the hypothesis was formulated that this problem could be
connected with properties of the decay amplitude
in the transition time region,
seems to be  possible manifestations of this effect and this
problem refers  to the first possibility mentioned above. There is
the problem with determining  the mass of broad resonances.
The measured range of possible mass of $\sigma$ meson is
very wide, 400 -- 1200 MeV. So one can not exclude the
possibility that the masses of some $\sigma$  mesons  are
measured for times of order  their lifetime and some of
them for times where their instantaneous energy
${\cal E}_{\sigma}(t)$ is much larger. This is exactly the
case presented in Fig. (\ref{h10}) and Fig. (\ref{h100}).
For broad mesons the ratio
$\frac{{\cal E}_{\sigma}^{0}}{\gamma_{\sigma}^{0}}$ is
relatively small and thus the time $t_{as}$ when the
above discussed effect occurs appears to be not too long.

Astrophysical and cosmological processes in which  extremely
huge numbers of unstable particles are created
seem to be  another possibility for the above discussed
effect to become manifest. The probability ${\cal P}_{\phi}(t)
= |a(t)|^{2}$ that an unstable particle, say $\phi$, survives
up to time $t \sim t_{as}$ is extremely small. Let
${\cal P}_{\phi}(t)$ be
\begin{equation}
{ {\cal P}_{\phi}(t)\,\vline }_{\;t \sim t_{as}}\;\sim\;10^{-k},
\label{p(t)-k}
\end{equation}
where $k \gg 1$, then there is a chance to observe some of
particles $\phi$ survived at $t \sim t_{as}$ only if there
is a source creating these particles in ${\cal N}_{\phi}$
number such that
\begin{equation}
{{\cal P}_{\phi}(t)\,\vline }_{\;t \sim
t_{as}}\;{\cal N}_{\phi} \; \gg \;1.
\label{N-phi}
\end{equation}
So if  a source exists that creates a flux containing
\begin{equation}
{\cal N}_{\phi} \;\sim\;10^{\,l},
\label{N-phi-l}
\end{equation}
unstable particles and $l \gg k$ then the probability
theory states that the number $N_{surv}$ unstable particles
\begin{equation}
N_{surv} = { {\cal P}_{\phi}(t)\,\vline }_{\;t \sim t_{as}}\;
{\cal N}_{\phi} \;\sim\;10^{l - k} \; \gg\;0,
\label{N-surv}
\end{equation}
has to survive up to time $t \sim t_{as}$. Sources creating
such numbers of unstable particles are known from cosmology
and astrophysics.
The Big Bang is the obvious example of such a source.
Some other examples include processes taking place in
galactic  nuclei (galactic cores) and inside  stars, etc.

\begin{figure}
\begin{center}
  \includegraphics[width=90mm]{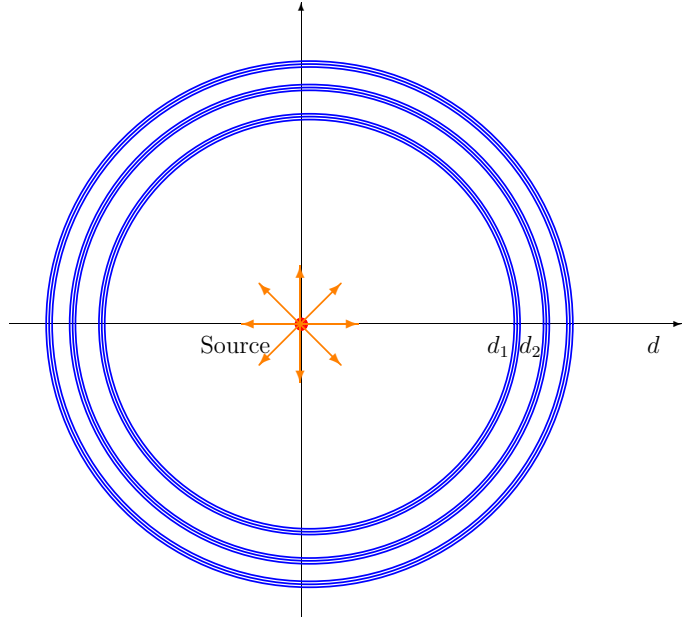}\\
  \end{center}
  \caption{Halos surrounding a source of unstable particles}
  \label{circles-a}
\end{figure}

So let us assume that we have an astrophysical source
creating a sufficiently large number of unstable particles
in unit of time and emitting a flux of these particles
and that this flux is constant or slowly varying in time.
Consider as an example a flux of neutrons. From (\ref{t-as-3})
it follows that for the neutron $t_{as}^{n}
\sim \,(250 \tau_{n} - \,300 \tau_{n})$, where
$\tau_{n} \simeq 886$ [s]. If the energies of these
neutrons are of order $30 \times 10^{17}$ [eV] then
during time $ t  \sim t_{as}^{n}$
they can reach a distance $d^{n} \sim 25 000$ [ly],
that is the distance of about a half of  the Milky Way
radius. Now if in a unit ot time a suitably large number
of neutrons ${\cal N}_{n}$ of the energies mentioned is
created by this source then in the distance $d^{n}$ from
the source
a number of spherically symmetric space areas (halos)
surrounding  the source, where neutron instantaneous
energies ${\cal E}_{n}(t)$ are much larger than  its
energy ${\cal E}_{n}^{0} =
\frac{m_{n}^{0} \,c^{2}}{\sqrt{1 - (\frac{v_{n}}{c})^{2}}}$,
($m_{n}^{0}$ is the neutron rest mass and $v_{n}$ denotes
its velocity) have to appear (see Fig. (\ref{circles-a})).
Of course this conclusion holds also for other unstable
particles $\phi_{\alpha}$ produced by this source.

Every kind of  particles $\phi_{\alpha}$ has its own halos
located at distances $d_{k}^{\phi_{\alpha}}$,
\[
d_{k}^{\phi_{\alpha}} \sim
v_{\phi_{\alpha}}\,t_{as}^{\phi,\,k},\;\;\;\;(k =1,2, \ldots),
\]
from the source. Radiuses  $d_{k}^{\phi_{\alpha}}$ of these
halos are determined by the particles'  velocities
$v_{\phi_{\alpha}}$ and by times  $t_{as}^{\phi,\,k}$
when instantaneous energies ${\cal E}_{\phi_{\alpha}}(t)$
have local maxima.

\begin{figure}
\begin{center}
  \includegraphics[width=80mm]{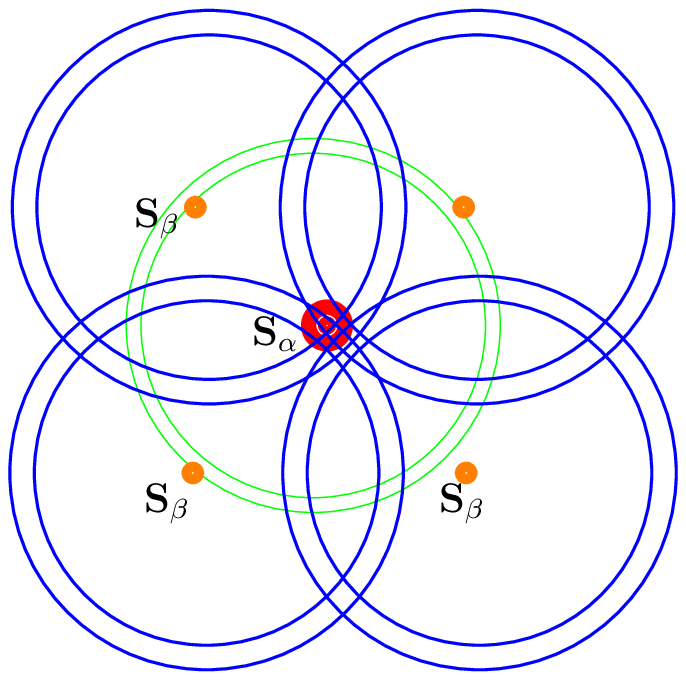}\\
  \end{center}
  \caption{Halos generated by a few not too
  distant  sources ${\bf S_{\alpha}}, {\bf S_{\beta}}$.}
  \label{circles-b}
\end{figure}

Unstable particles $\phi_{\alpha}$  forming these halos
and having instantaneous energies   ${\cal E}_{\phi_{\alpha}}(t)
\gg {\cal E}_{\phi_{\alpha}}^{0} = m_{\phi_{\alpha}}^{0}\,c^{2}$
have to interact gravitationally with objects outside of these
halos as particles of masses $m_{\phi_{\alpha}}(t) =
\frac{1}{c^{2}}\,   {\cal E}_{\phi_{\alpha}}(t) \,
\gg\, m_{\phi_{\alpha}}^{0}$. The possible observable
effects depend on the astrophysical source of these particles considered.

If the halos are formed by unstable particles emitted
as a result of internal star processes then
in the case of very young stars cosmic dust and gases
should be attracted by these halos as a result of a gravity
attraction. So, the halos should be a places where the
dust and gases condensate. Thus in the case of very young
stars one may consider the halos as the places  where
planets are born. On the other hand in the case of much
older stars a presence of halos should manifest itself
in tiny changes of velocities and accelerations of object
moving in  the considered planetary star system  relating
to those calculated without taking into account of
the halos presence.

If the halos are formed by unstable particles emitted
by a galaxy core and these particles are such that
the ratio $\frac{{\cal E}_{\phi}(t)}{{\cal E}_{\phi}^{0}}$
is suitably large inside the halos,
then rotational velocities of stars rounding the galaxy center
outside the halos should differ from those calculated without
taking into account the halos. Thus the halos may affect the
form of rotation curves of galaxies. (Of course, we do do not
assume that the   sole factor affecting  the form of the
rotation curves are these halos). Another possible
effect is that the
velocities of particles crossing these galactic  halos
should slightly vary in time due to gravitational
interactions, i. e. they should gain some acceleration.
This should cause charged particles to emit
electromagnetic radiation when  they cross the halo.

Note that the above mentioned effects seems to be
possible to examine. All these effect are the simple
consequence of the fact that the instantaneous energy
${\cal E}_{\phi_{\alpha}}(t)$ of unstable particles becomes
large  compared with ${\cal E}_{\phi_{\alpha}}^{0}$ and for
some times even extremely large. On the other hand this
property of ${\cal E}_{\phi_{\alpha}}(t)$ results from
the rigorous analysis of properties of the quantum
mechanical survival probability $a(t)$ (see (\ref{a(t)}) )
and from the assumption that the energy spectrum is
 bounded from below.

\end{document}